\documentclass[epj]{svjour}
\usepackage{amsmath}
\usepackage{amssymb}
\usepackage{epsfig}
\def\bea{\begin{eqnarray}}
\def\beann{\begin{eqnarray*}}
\def\beq{\begin{equation}}
\def\eea{\end{eqnarray}}
\def\eeann{\end{eqnarray*}}
\def\eeq{\end{equation}}
\def\nn{\nonumber}
\def\xp{\vec{p}}
\def\xpp{\vec{p}\,'}
\begin{document}
\title{{\boldmath$\bar{D}N$} interaction from meson-exchange and quark-gluon dynamics}
\author{J. Haidenbauer\inst{1}, G. Krein\inst{2}, Ulf-G. Mei{\ss}ner\inst{1,3}, 
and A. Sibirtsev\inst{3}}

\institute{
Forschungszentrum J\"ulich, Institut f\"ur Kernphysik,
D-52425 J\"ulich, Germany \and
Instituto de F\'{\i}sica Te\'{o}rica, Universidade Estadual
Paulista,
Rua Pamplona, 145 - 01405-900 S\~{a}o Paulo, SP, Brazil
\and
Helmholtz-Institut f\"ur Strahlen- und Kernphysik (Theorie),
Universit\"at Bonn, Nu\ss allee 14-16, D-53115 Bonn, Germany
}
\date{Received: date / Revised version: date}
 
\abstract{
We investigate the $\bar D N$ interaction at low energies using a meson-exchange 
model supplemented with a short-distance contribution from one-gluon-exchange. 
The model is developed in close analogy to the meson-exchange $KN$ interaction 
of the J\"ulich group utilizing SU(4) symmetry constraints. The main 
ingredients of the interaction are provided by vector meson ($\rho$, $\omega$)
exchange and higher-order box diagrams involving ${\bar D}^*N$, 
$\bar D \Delta$, and ${\bar D}^*\Delta$ intermediate states. The short 
range part is assumed to receive additional contributions from genuine 
quark-gluon processes. The predicted cross sections for $\bar D N$ for 
excess energies up to 150 MeV are of the same order of magnitude as those 
for $KN$ but with average values of around 20 mb, roughly a factor two 
larger than for the latter system. It is found that the $\omega$-exchange 
plays a very important role. Its interference pattern with the
$\rho$-exchange,  which is basically fixed by the assumed SU(4) symmetry, clearly 
determines the qualitative features of the $\bar D N$ interaction -- very 
similiar to what happens also for the $KN$ system. 
}
 

\authorrunning{J. Haidenbauer et al.}
\titlerunning{$\bar DN$ interaction}

\maketitle

\section{Introduction}

The study of the interactions of charmed hadrons with nucleons is
of interest in several contexts. One example is in experiments of 
relativistic heavy ion collisions (RHIC). Since long 
time~\cite{Matsui:1986dk} the suppression of $J/\Psi$ production in RHIC 
is being considered as a possible signature for the formation of a 
quark-gluon plasma (QGP). The alleged suppression would occur
because the deconfined quarks of the QGP screen the long range
confining potential thus making impossible the formation of the
mesonic bound states. However, collisions of the charmed mesons
with hadrons in the medium can also lead to dissociation of these
mesons, subverting therefore the screening scenario. Moreover,
more recently it has been argued that heavy quarkonia could be
re-formed via rescattering processes of open-charm hadrons in the
late stages of RHIC which would then lead to an enhanced 
$J/\Psi$ production \cite{Thews}. 
Thus, it seems clear that a good knowledge of the interaction of
charmed mesons with ordinary hadrons like nucleons is a prerequisite
for differentiating between these scenarios. For a recent review on 
these issues, see Ref.~\cite{Satz:2005hx}. 
Another example where 
the interaction of charmed mesons with ordinary hadrons is of interest
refers to studies of chiral symmetry restoration in a hot and/or
dense medium. In this respect, the interaction of charmed $D$
mesons - which are composed of one light and one heavy quark -
with nucleons is of particular interest. The properties of the
light quarks in a $D$ meson are sensitive to temperature and
density and, therefore, changes in the properties of the $D$ mesons
in medium can be expected. Consequently, one can also expect that
their interactions with nucleons will change in the medium. The
$D$ meson and its lowest excitations are somewhat special in this respect
because their spectroscopy is simpler than of ordinary mesons
composed solely by $u$ and $d$ quarks. This is so because the
charm quark $c$ is much heavier than the light $u$ and $d$ quarks,
and to a good approximation these mesons can be described as
one-body bound states, a fact that simplifies tremendously 
their~study. 

Before one can infer in a sensible way changes of the interaction 
in the medium, a reasonable understanding of the interaction 
in free space is required. However, here one has to cope with a
major difficulty, namely the complete lack of experimental data at low 
energies for the free-space interaction. 
This situation is hopefully going to change soon with the 
operation of the FAIR facility at the GSI laboratory in Germany. 
There are proposals for experiments by the $\bar{\rm P}$ANDA 
collaboration~\cite{Panda} at this facility 
to produce $D$ mesons by annihilating antiprotons on the deuteron and, 
through the rescattering of the produced $D$ and $\bar D$ mesons on the spectator 
nucleon~\cite{Cassing:1999wp}, to determine $DN$ as well as $\bar D N$ cross 
sections and possibly 
even phase shifts. Still, for the design of detectors and of efficient data 
acquisition systems, estimates for the magnitude of such cross sections are 
urgently required.
Therefore, there is a need for developing models of the interaction
of charmed particles with ordinary hadrons, 
and since -- as said -- not much is known empirically, 
such models can only and should be constrained as much as 
possible by symmetry arguments, analogies with other similar 
processes, and the use of different degrees of freedom. 

The interaction of charmed mesons with ordinary had{\-}rons composed of $u$ and 
$d$ quarks has been investigated using effective hadronic Lagrangians and 
quark models. Most of the studies have concentrated on the interaction of 
$J/\Psi$ and other heavy charmonia with ordinary hadrons, mainly due to the 
interest in the QGP suppression hypothesis alluded above -- see 
Ref.~\cite{Barnes:2003vt} for a review on these investigations. With 
respect to the interaction of the $D$ meson with the nucleon, which is the 
subject of the present paper, not much is known. The work of 
Ref.~\cite{Lin00}, using an effective SU(4) 
hadronic Lagrangian, to the best of our knowledge was the first one to 
provide estimates of cross sections for the $DN$ system in Born approximation. 
In terms of quark degrees of freedom, the authors of 
Ref.~\cite{Sibirtsev:1999js} have made an estimate for the $DN$ cross sections 
using quark rearrangement arguments and concluded that such cross sections 
should be equal to the corresponding $KN$ cross sections, though no explicit 
model was employed. (See also the results presented in Ref.~\cite{Sib01}).

In the present paper we investigate the $\bar DN$ interaction within a 
meson-exchange model and a quark model utilizing one-gluon-exchange 
(OGE), in the spirit of a recent study of the $KN$ system by 
us~\cite{HHK}. ($\bar D$ is used here generically for the $\bar D^0$ and 
$D^-$ isospin doublet which contains a $\bar c$ quark, and corresponds 
to $K$ consisting of the $K^+$ and $K^0$ isospin doublet with an $\bar s$ quark.)
To be more specific, the $\bar DN$ interaction 
we construct is an extension of the $KN$ meson-exchange model of the 
J\"ulich group~\cite{Juel1,Juel2,Juel3}, generalized by assuming as a working
hypothesis SU(4) symmetry constraints. 
Note that the $KN$ model described in Refs.~\cite{Juel1,Juel2} considered not
only single boson exchanges ($\sigma$, $\rho$, $\omega$), but also contributions 
from higher-order diagrams involving $N$, $\Delta$, $K$ and $K^*$ intermediate 
states. 
We focus on the $\bar DN$ system because it has the advantage that its
dynamics should be governed predominantly by the same ``long-range'' physics 
as the $KN$ interaction, i.e. by the exchange of ordinary (vector and 
possibly scalar) mesons. Thus, fairly reliable and, most importantly, 
essentially parameter-free predictions can be made once one accepts the
constraints provided by SU(4) symmetry. The $DN$ system is expected to 
exhibit a much richer structure \cite{Hofmann05} and thus may be more interesting 
\cite{Tolos04,Lutz05,Mizutani06}, but it involves also much larger uncertainties. In this 
case, like in the analogous $\bar KN$ system, there are couplings to several other 
channels which are already open near the $DN$ threshold ($\Lambda_c \pi$, $\Sigma_c\pi$)  
or open not far from the threshold ($\Lambda_c \eta$). 
It is obvious that the coupling to those channels must play a crucial role for the 
dynamics of the $DN$ system -- as it does in the corresponding $\bar KN$ system --
and, thus, will have a strong impact on any quantitative results.
But the transitions to those channels and the interactions
in those channels involve charmed baryon resonances as well as
the exchange of charmed mesons, for example the $D^*$(2010),
whose coupling constants and associated vertex form factors, required 
in any meson-exchange model, are practically unknown and difficult to
constrain. 

Our $\bar{D}N$ model is obtained by substituting the 
one-boson-exchange contributions, but also the box diagrams involving 
$K^*N$, $K\Delta$, and $K^*\Delta$ intermediate states,
of the original $KN$ model of the J\"ulich group
by the corresponding contributions to the $\bar{D}N$ 
interaction under the constraint of SU(4) symmetry. 
Regarding the short-ranged quark
part, we use the dominat contributions of OGE exchange, which are the
Coulomb and spin-spin parts. These are of the same form as for the $KN$ 
interaction, but the mass of the charm  quark is much heavier than the one
of the strange quark, and the size parameter of the meson wave function 
is also different. Therefore, the $\bar DN$ interaction will be different from the 
$KN$ interaction. We want to emphasize that we iterate the effective meson-exchange 
and OGE potentials in a Lippmann-Schwinger equation, contrary to the common 
practice of using only Born approximation \cite{Lin00,Sibirtsev:1999js,Sib01}. 
We found that the effect of iteration on the predicted cross sections 
can be quite substantial, being of the order of 50 \% in some cases. 

The plan of our paper is the following. In the next section we specify the 
three-meson vertices used in the paper. Specifically, we discuss the constraints of
SU(4) symmetry to relate the required new couplings and present the numerical
values of vertex parameters. The interaction Lagrangians, which are needed to 
complete the derivation of the meson-baryon potential, are presented in 
Appendix~\ref{app:lags}. In Section~\ref{sec:DN-qex} we discuss the quark-gluon 
exchange mechanism for the $\bar DN$ interaction. The detailed derivation of the 
equations shown in this Section are outlined in Appedix~\ref{app:qm}. Our 
numerial results are presented in Section~\ref{sec:res}. 
The paper ends with a short summary.

\section{The ${\bf \bar D N}$ interaction in the meson-exchange picture}
\label{sec:DN-mex}

\begin{figure}[t]
\vspace*{+1mm}
\centerline{\hspace*{3mm}
\psfig{file=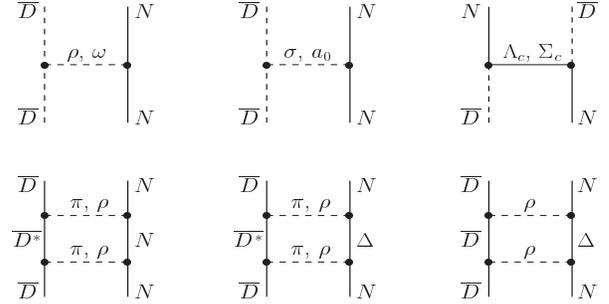,width=11.0cm,height=13.0cm}}
\vspace*{-7.5cm}
\caption{Meson-exchange contributions included in the $\bar D N$
interaction.
}
\label{Diag}
\end{figure}

The meson-exchange model of the $\bar{D}N$ interaction is constructed
in close analogy to the corresponding $KN$ potentials developed by
the J\"ulich group some time ago \cite{Juel1,Juel2}. In those models,
derived within time-ordered perturbation theory,
not only single-meson (and baryon) exchanges were taken into account,
but also higher-order box diagrams involving $K^*N$, $K\Delta$, and
$K^*\Delta$ intermediate states. Thus, we will consider the
corresponding contributions to the $\bar{D}N$ interaction too,
cf. Fig.~\ref{Diag}.
The general scheme and also the explicit expressions for the
various contributions to the interaction potential are described
in detail in Refs.~\cite{Juel1,Juel2} and, therefore, we do not reproduce 
them in the present paper. We only summarize the used interaction Lagrangians
in Appendix A. Here we want to focus on the SU(4) structure
which is used to extend and relate the $\bar{D}N$ interaction to the
$KN$ system.

\begin{table*}[t]
\caption{SU(4) matrix representation of the pseudo-scalar ($P$) and vector ($V$)
mesons.}
\label{SU4}
\begin{eqnarray} 
P&=&
\left (
\begin{array}{cccc}
\frac{\pi^0}{\sqrt 2}+\frac{\eta}{\sqrt 6}+\frac{\eta_c}{\sqrt{12}}
& \pi^+ & K^+ & {\bar D}^0 \\
\pi^- & -\frac{\pi^0}{\sqrt 2}+\frac{\eta}{\sqrt 6}+\frac{\eta_c}{\sqrt{12}}
& K^0 & D^- \\
K^- & {\bar K}^0 & -\sqrt {\frac{2}{3}}\eta+\frac{\eta_c}{\sqrt{12}}
& D_s^- \\
D^0 & D^+ & D_s^+ & -\frac{3\eta_c}{\sqrt{12}}
\end{array}
\right )  \nonumber \\[2ex]
V&=&
\left (
\begin{array}{cccc}
\frac{\rho^0}{\sqrt 2}+\frac{\omega_8}{\sqrt 6}+\frac{\omega_{15}}{\sqrt {12}}
& \rho^+ & K^{*+} & {\bar D}^{*0} \\
\rho^- & -\frac{\rho^0}{\sqrt 2}+\frac{\omega_8}{\sqrt 6}+\frac{\omega_{15}}
{\sqrt {12}} & K^{*0} & D^{*-} \\
K^{*-} & {\bar K}^{*0} & -\sqrt {\frac{2}{3}}\omega_8+\frac{\omega_{15}}{\sqrt {12}}
& D_s^{*-} \\
D^{*0} & D^{*+} & D_s^{*+} & -\frac{3\omega_{15}}{\sqrt {12}}
\end{array}
\right )  \nonumber
\end{eqnarray}
\end{table*}
For the construction of the $\bar{D}N$ interaction we need
three-meson vertices involving charmed mesons of the kind $PPV$
and $VVP$ ($P$ = pseudoscalar meson, $V$ = vector meson). The
general form of the SU(4) invariant Lagrangian~is
\begin{eqnarray}
\nonumber
{\cal L}_{MMM} &=& g_{\{15\}} [-\alpha Tr([M_{\{15\}},M_{\{15\}}]M_{\{15\}}) \\
\nonumber
&+& (1-\alpha) Tr(\{M_{\{15\}},M_{\{15\}}\}M_{\{15\}}) ] \\ 
\nonumber
&+& g_{\{15\}\{15\}\{1\}} (1-\alpha) Tr(\{M_{\{15\}},M_{\{15\}}\}M_{\{1\}}) \\
\nonumber
&+& g_{\{15\}\{1\}\{15\}}(1-\alpha) Tr(\{M_{\{15\}},M_{\{1\}}\}M_{\{15\}}) \\
&+& g_{\{1\}}(1-\alpha) Tr(\{M_{\{1\}},M_{\{1\}}\}M_{\{1\}}) \ ,
\label{lsu4}
\end{eqnarray}
where $\alpha$ is the $F/(F+D)$ ratio and $M_{\{15\}}$ ($M_{\{1\}}$) stands for the
SU(4) meson--15-plet (-singlet) matrix. For pseudo-scalar ($P$) and vector ($V$)
mesons $M_{\{15\}}$ are $4\times4$ matrices which are given in Table \ref{SU4}. 
Note that the $PPV$ vertices involve only $F$-type coupling ($\alpha = 1$)
if we require charge conjugation invariance while the $VVP$ vertices involve
only $D$-type coupling ($\alpha = 0$). Let us stress again that SU(4) should
not be a flavor symmetry of QCD but rather a working hypothesis to get a
handle on the various couplings and form factors employed in our model. It is,
of course, strongly broken due to the use of the very different physical
masses of the  various mesons.

Based on the assumed SU(4) symmetry all relevant three-meson coupling constants
can be derived from the empirically known $\pi\pi\rho$ coupling. In the
J\"ulich model \cite{Juel2} the value $g_{\pi\pi\rho} = 6.0$ is used. The
coupling constants of the other vertices that follow from this value
are listed in Table \ref{coup}.

\begin{table*}[ht]
\caption{Vertex parameters used in the meson-exchange model of the
$\bar DN$ interaction at the $\bar D\bar D m_r$ or $\bar D \bar
D^* m_r$ (M) and $NNm_r$ or $N\Delta m_r$ (B) vertices. $M_r$ and
$m_r$ refers to the mass of the exchanged particle. Note that the
scalar meson exchanges ($\sigma$, $a_0$) are considered as an
effective interaction, cf. text, and therefore we provide only the
product of the coupling constants. }
\begin{center}
\begin{tabular}{|c|cccccc|}
\hline\noalign{\smallskip}
Process & Exch. part. & $M_r$ or $m_r$
 & $g_M /\sqrt{4 \pi}$ & $g_B /\sqrt{4 \pi}$ [$f_B/g_B$]
 & $\Lambda_M$  & $\Lambda_B$   \\
        &  & [MeV]   &  & & \ [GeV] \ & \ [GeV] \ \\
\hline\noalign{\smallskip} 
$\bar D N \rightarrow \bar D N$
                     & $\rho$         & \phantom{1}769\phantom{.03}
 &\phantom{--4}0.843 &\phantom{--4}0.917 [6.1]
 & 1.4 & 1.6  \\
                     & $\omega$       & \phantom{1}782.6\phantom{3}
 &\phantom{--4}0.843  &\phantom{--4}2.750 [0.0]
 & 1.5 & 1.5  \\
                     & $\Lambda_c$      & 2285\phantom{.03}
 &\phantom{-4}-2.284 &\phantom{-4}-2.284\phantom{[6.1]}
 & 4.1 & 4.1  \\
                     & $\Sigma_c$       & 2455\phantom{.03}
 &\phantom{--4}0.435 &\phantom{--4}0.435\phantom{[6.1]}
 & 4.1 & 4.1  \\
 & $\sigma$       & \phantom{1}600\phantom{.03}
 &\multicolumn{2}{c}{0.25 (1.00)} & 1.7 & 1.2  \\
 & $a_0$       & \phantom{1}980\phantom{.03}
 &\multicolumn{2}{c}{0.65 (2.60)} & 1.5 & 1.5  \\
$\bar D N \rightarrow {\bar D}^* N$ & $\pi$          &
\phantom{1}138.03
 &\phantom{--4}0.843 &\phantom{--4}3.795\phantom{[6.1]}
 & 1.3 & 0.8  \\
                        & $\rho$         & \phantom{1}769\phantom{.03}
 &\phantom{--4}0.843  &\phantom{--4}0.917 [6.1]
 & 1.4 & 1.0  \\
$\bar D N \rightarrow {\bar D}^* \Delta$ & $\pi$          &
\phantom{1}138.03
 &\phantom{--4}0.843 &\phantom{--4}0.600\phantom{[6.1]}
 & 1.2 & 0.8  \\
                          & $\rho$         & \phantom{1}769\phantom{.03}
 &\phantom{--4}0.843 &\phantom{--4}5.740\phantom{[6.1]}
 & 1.3 & 1.0  \\
$\bar D N \rightarrow \bar D  \Delta$ & $\rho$         &
\phantom{1}769\phantom{.03}
 &\phantom{--4}0.843 &\phantom{--4}5.470\phantom{[6.1]}
 & 1.3 & 1.6 \\
\hline
\end{tabular}
\end{center}
\label{coup}
\end{table*}

As far as the coupling constants belonging to the $NN$ and $N\Delta$ vertices
are concerned we take precisely the same values as in Ref. \cite{Juel2},
which are based on those of the (full) Bonn $NN$ potential, cf. Ref.~\cite{MHE}.
These coupling constants are listed in Table \ref{coup} too.
The J\"ulich $KN$ potential contains also vertex form factors $F$
that are meant to take into account the extended hadron structure
and are parametrized in the conventional monopole or dipole form
\cite{Juel1,Juel2}. In the present study of the $\bar DN$ system
the cut-off masses appearing in those form factors for the various
three-meson and baryon-baryon-meson vertices are likewise taken
over from Ref.~\cite{Juel2}. Specifically, we make the assumption
that $F_{\bar D\bar D m}(\vec q_m^{\, 2}) \simeq F_{KKm}(\vec
q_m^{\, 2})$. This prescription is motivated by the notion that
those form factors parametrize predominantly the off-mass-shell
behaviour of the exchanged particles -- which are indeed the same 
in the $KN$ and in the $\bar D N$ interaction.

Let us make some more comments about the coupling constants at the
three-meson vertices.
$SU(4)$ symmetry implies the following for the vector meson coupling
constants relevant for our study:
\begin{eqnarray}
\nonumber
g_{KK\omega_8}&=&\sqrt{3} g_{KK\rho } = \sqrt{3}\, \frac{1}{2}\, g_{\pi\pi\rho} , \ \
g_{KK\omega_{15}}=0 \\
g_{\bar D \bar D\omega_8}&=&\sqrt{\frac{1}{3}} g_{KK\rho} , \ \
g_{\bar D \bar D\omega_{15}}=\sqrt{\frac{8}{3}} g_{KK\rho}
\label{omega}
\end{eqnarray}
\begin{eqnarray}
g_{\bar D \bar D\rho } =g_{KK\rho } = \frac {g_{\pi\pi\rho}}{2} \ .
\label{rho}
\end{eqnarray}
Assuming ideal mixing of the $\omega_{15}$, $\omega_8$ and $\omega_1$ one obtains
for the coupling constants of the physical $\omega$ and $\phi$
\begin{eqnarray}
\nonumber
g_{\bar D\bar D\omega}&=&\sqrt{\frac{1}{2}}g_{\bar D\bar D\omega_1}+
\sqrt{\frac{1}{3}}g_{\bar D\bar D\omega_8} +
\sqrt{\frac{1}{6}}g_{\bar D\bar D\omega_{15}} \\
g_{\bar D\bar D\phi}&=&-\sqrt{\frac{1}{4}}g_{\bar D\bar D\omega_1}+
\sqrt{\frac{2}{3}}g_{\bar D\bar D\omega_8} -
\sqrt{\frac{1}{12}}g_{\bar D\bar D\omega_{15}} \ .
\end{eqnarray}
The same relation holds also for the $K$ meson. In case of the $K$ meson
the coupling constant $g_{KK\omega}$ is given by that of $g_{KK\omega_8}$ alone,
since there is no singlet coupling for $PPV$ vertices as mentioned above:
\begin{eqnarray}
g_{KK\omega}&=&\sqrt{\frac{1}{3}} g_{KK\omega_8} = g_{KK\rho} \ .
\label{omegaK}
\end{eqnarray}
This is the coupling constant used in the J\"ulich $KN$ models
\cite{Juel1,Juel2}.
In case of the $D$ meson the coupling constant is given by
\begin{eqnarray}
g_{\bar D\bar D\omega}&=&\sqrt{\frac{1}{3}}g_{\bar D\bar D\omega_8} +
\sqrt{\frac{1}{6}}g_{\bar D\bar D\omega_{15}}
= g_{KK\rho} \ .
\label{omegaD}
\end{eqnarray}
Summarizing the above results from Eqs.~(\ref{rho},\ref{omegaK},\ref{omegaD})
we see that
\begin{eqnarray}
\nonumber
g_{\bar D\bar D\rho}&=&g_{KK\rho} \\
g_{\bar D\bar D\omega}&=&g_{KK\omega} \ .
\label{Coupling}
\end{eqnarray}
Thus, the coupling constants of the exchanged vector me\-sons are
the same for the $KN$ and $\bar DN$ systems under assumption of $SU(4)$ symmetry and
ideal mixing.

Since some of the couplings involving
the $D$ meson are known empirically, at least to some extent, we want
to review them briefly here. The $DD\rho$ coupling constant was determined
in Refs. \cite{Mat98,Lin00a} based on the vector dominance model and found to be
$g_{DD\rho} = 2.52-2.8$. This value, which was subsequently adopted in several
investigations~\cite{Lin00,Lin01,Liu02}, is only marginally smaller than
the one which follows from assuming SU(4) symmetry. The same is true for the
$DD\omega$ coupling constant, found to be $g_{DD\omega} = -2.84$ in 
Ref. \cite{Lin00a}, likewise derived within the vector dominance model. 
In Ref. \cite{Liu02} the value $g_{\pi D D^*} = 5.56$ is cited, derived
from the measured decay width of the $D^*$ meson. Here the corresponding
SU(4) coupling constant is roughly a factor 2 smaller.

In any case, in our model calculation we use coupling constants that
are determined fully by SU(4) symmetry. The difference to those values
deduced from available experimental information is not very large and,
thus, does not really warrant a departure from SU(4) at present. 
Indeed, there are other assumptions made in the model calculation, 
that could be considered to be more questionable, for
example those about the vertex form factor. As mentioned 
above, the prescription we use relies on the fact that the same particles
are exchanged in the $KN$ and $\bar D N$ potentials. Possible influences
from differences in the off-mass-shell dependence due to the different 
($KN$ or $\bar D N$) intermediate states, appearing in higher iterations, 
are simply ignored.
However, the main uncertainty in the meson-exchange model arises from
the treatment of the scalar-meson sector. Here, unlike for pseudoscalar
and vector mesons, so far there is no general agreement
about who are the actual members of the lowest lying scalar-meson SU(3)
multiplet. (For a thorough discussion on that issue and an overview of the
extensive literature we refer the reader to \cite{Kle04,Kal05} and
references therein.)
Therefore, it remains unclear whether and how the relations for the coupling
constants given in Eq. (\ref{lsu4}) should be applied in the SU(3) case
\cite{Juel2,Hai05}, but even more so when it comes to SU(4).
It is known for a long time that the contributions from the scalar sector
play a crucial role in any baryon-baryon and meson-baryon interaction at
intermediate ranges.

In the present paper we consider two different scenarios for the scalar
mesons. First, in line with the works in Refs.~\cite{Juel2,Hai05}, we view
the contributions in the scalar sector as being due to correlated $\pi\pi$
and $K\bar K$ exchange. However, in the absence of a concrete model
for those contributions, as it was used in \cite{Juel2,Hai05}, we resort
here to a rough estimation of their strength. Based on the scale of
the correlated $\pi\pi$ and $K\bar K$ exchange, which we identify with
the masses of the relevant propagators -- $\rho$ ($K^*$) and/or $\pi\pi$ (
$K\bar K$) for the $\pi N$ and $KN$ interactions, but the $D^*$(2010)
and/or $\bar DD$ for the $\bar D N$ system -- we expect that its
strength should be about 4 times smaller in the latter case. Thus,
we reduce the coupling constants by that factor as compared to the
values used in \cite{HHK}. Note that this reduction is supported by
available model calculations of the reaction $p \bar p \to D \bar D$
\cite{Kro89,Kai94} which suggest that the corresponding amplitude,
which would form the main ingredient for a microscopic calculation
of the correlated $\pi\pi$ exchange for $\bar D N$, cf. Ref. \cite{Juel2},
is significantly smaller than the one for $p \bar p \to K \bar K$,
even when taking into account the kinematical differences. The used
values are given in Table \ref{coup}. 
The second scenario is an attempt to simulate the case that the scalar 
contributions are due to genuine scalar-meson exchange. 
Accordingly, we use the same scalar coupling constants 
for the $\bar DN$ interaction as in the $KN$ model \cite{HHK}. 
The concrete values for that scenario are listed in brackets in 
Table \ref{coup}.

\section{The ${\bf \bar DN}$ interaction based on the quark-gluon exchange mechanism}
\label{sec:DN-qex}

\begin{figure}[ht]
\begin{center}
\includegraphics[width=6.cm]{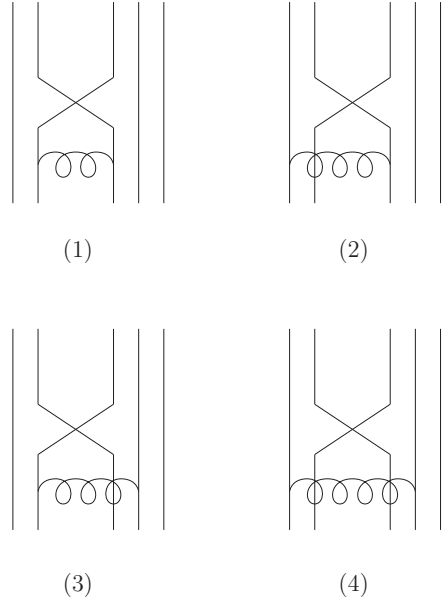}
\caption{Pictorial representation of the four different
quark-interchange processes that contribute to the $\bar DN$
interaction. The wavy lines represent the one-gluon exchange (OGE)
and the solid lines represent quarks. } \label{fig:interch}
\end{center}
\end{figure}

The quark interchange processes with one-gluon-exchange (OGE) we consider 
in the present paper are represented pictorially in Fig.~\ref{fig:interch}. 
In these graphs, the $D$ mesons are $\bar{D}^0 = u\,\bar{c}$ and $D^{-} =
d\,\bar{c}$, so that the exchanged quarks are always the light $u$
and $d$ quarks, the $\bar{c}$ antiquarks are not interchanged. The dominant 
contributions of the OGE interaction are the Coulomb and spin-spin parts. The
interaction of the quarks $i$ and $j$ with constituent masses $m_i$ and $m_j$ 
can be written as $V_{ij} = T^a_i T^a_j \, V_{ij}(q,S)$, where 
$T^a = \lambda^a/2$ for a quark and $T^a = - (\lambda^a)^T/2$ for an 
antiquark, and the momentum and spin dependent pieces as $V_{ij}(q,S)$ 
are given as
\bea V_{ij}(q,S) &=& v_{C}(q) + v_{SS}(q)\, 
\vec{S}_i{\cdot}\vec{S}_j \nn \\ &=& \frac{4\pi\alpha_s}{q^2} -
\frac{8\pi\alpha_s}{3m_im_j}\,\vec{S}_i{\cdot}\vec{S}_j ,
\label{VqS} \eea
where $\alpha_s$ is the quark-gluon coupling constant.

As shown in more detail in Appendix~\ref{app:qm}, the effective
$\bar DN$ potential ${\cal V}_{\bar DN}$ can be written as a sum of four
contributions~as
\beq {\cal V}_{\bar DN}(\xp, \xpp) = \sum^4_{i=1} \omega_i \, \left[V_i(\vec{p},
\vec{p}\,') +V_i(\vec{p}\,', \vec{p})\right]/2, \eeq
where each term in the sum corresponds to a graph in
Fig.~\ref{fig:interch}. The $\omega_i$'s are given
Table~\ref{tab:wi}; they come from summing over the color-flavor-spin
indices of the quarks and include symmetry combinatorial factors. The
$V_i$'s are functions of the center-of-mass momenta $\xp$ and
$\xpp$ which are given by multidimensional overlap integrals over
the internal wave functions of the nucleons and mesons and the OGE
potentials. When using Gaussian forms for the nucleon and meson
wave functions, many of the integrals can be done analytically and
the $V_i$'s can then be expressed in terms of a single
three-dimensional integral as
\bea 
\nonumber 
V_i(\xp,\xpp) =&& e^{- a_i p^2 \, - \, b_i {p'}^{\,2} + c_i
\xp{\cdot}\xpp} \left[\frac{3g}{(3+2g)\pi\alpha^2_N}\right]^{3/2} \\
&&\times \int \frac{d^3 q}{(2\pi)^3} \, v(q) \, e^{- d_i q^2 + {\vec{e}}_i
{\cdot}\vec{q} } \ , \label{Vi} 
\eea
where $v(q) = v_C(q)$, or $v_{SS}(q)$ defined in Eq.~(\ref{VqS}), 
$g=\alpha^2_N/\beta^2_D$, where $\alpha_N$ and $\beta_D$ are the Gaussian 
widths of the nucleon and the $D$ meson. The $a_i, b_i, \cdots$ 
are given in terms of $g$ and the quark masses -- see Appendix~\ref{app:qm}.

\begin{table*}[ht]
\caption{The color-spin-flavor coefficients from the spin-spin OGE
interaction for the $\omega_i$ for the $D^-N$ and $\bar{D}^0N$
systems for the individual charge states, and for the $I=0$ and
$I=1$ combined isospin states.}
\begin{center}
\begin{tabular}{|c|cccc|}
\hline\noalign{\smallskip}
Process & $\omega_1$ & $\omega_2$  & $\omega_3$ & $\omega_4$    \\
{} &$1_i 1_j$ \hspace{0.25cm} $S_i S_j$ &$1_i 1_j$ \hspace{0.25cm}
$S_i S_j$ &$1_i 1_j$ \hspace{0.25cm} $S_i S_j$ &$1_i 1_j$ \hspace{0.25cm} $S_i S_j$ \\
\hline\noalign{\smallskip} 
$p\,\bar{D}^0 \rightarrow p\,\bar{D}^0$ & 1/3 \hspace{0.25cm} 1/3 &
1/3 \hspace{0.25cm} 1/3
& 1/3 \hspace{0.25cm} 1/18 & 1/3 \hspace{0.25cm} 1/18 \\
$n\,D^- \rightarrow  n\,D^-$ & 1/3 \hspace{0.25cm} 1/3 & 1/3
\hspace{0.25cm} 1/3
& 1/3 \hspace{0.25cm} 1/18 & 1/3 \hspace{0.25cm} 1/18 \\
$p\,D^- \rightarrow  p\,D^-$ & 1/3 \hspace{0.25cm} 1/6 & 1/3
\hspace{0.25cm} 1/6
& 1/3 \hspace{0.25cm} 1/9 & 1/3 \hspace{0.25cm} 1/9 \\
$n\,\bar{D}^0 \rightarrow n\,\bar{D}^0$ & 1/3 \hspace{0.25cm} 1/6
& 1/3 \hspace{0.25cm} 1/6
& 1/3 \hspace{0.25cm} 1/9 & 1/3 \hspace{0.25cm} 1/9  \\
$p\,D^- \rightarrow n\,\bar{D}^0$ & 1/3 \hspace{0.25cm} 1/6 & 1/3
\hspace{0.25cm} 1/6
& 1/3 \hspace{0.25cm} -1/18 & 1/3 \hspace{0.25cm} -1/18  \\
\hline
$I=0$ & 0 \hspace{0.25cm} 0  &  0 \hspace{0.25cm} 0
& 0 \hspace{0.25cm} - 1/6  & 0 \hspace{0.25cm} - 1/6 \\
$I=1$ & - 4/9 \hspace{0.25cm} - 1/3 & + 4/9 \hspace{0.25cm} - 1/3
& + 4/9 \hspace{0.25cm} - 1/18 & -4/9 \hspace{0.25cm} - 1/18 \\
\hline
\end{tabular}
\end{center}
\label{tab:wi}
\end{table*}

\section{Results and discussion}
\label{sec:res}

The original $KN$ model of the J\"ulich group includes besides
the exchange of the standard mesons also an
additional phenomenological (extremely short-ranged) repulsive
contribution, a ``$\sigma_{rep}$'', with a mass of about 1.2
Gev \cite{Juel2}. This contribution was introduced ad-hoc in
order to achieve a simultaneous description of the empirical $KN$
$S$- and $P$-wave phase shifts, but it is also required for a
consistent description of the $KN$ and $\bar KN$ systems \cite{Mueller},
which are interrelated via a $G$-Parity transformation.
Evidently, due to its phenomenological nature it remains unclear
how that contribution should be treated when going over to the
$\bar DN$ system.
Fortunately, a recent investigation by our group provided evidence that
a significant part of that short-ranged repulsion required in the
original J\"ulich model could be due to genuine quark-gluon
exchange processes. Thus, in the present study we will build upon this
insight when constructing a model of the $\bar DN$ interaction.
In particular this means that also for the $\bar DN$ system we consider
contributions from meson-exchange as well as from quark-gluon mechanisms
where each of them is closely linked to the corresponding pieces in
the $KN$ interaction.
However, in order to get a better understanding on what changes and what
remains the same in the transition from $KN$ to $\bar DN$ we will first
study those sectors separately. In order to facilitate an easy comparison
of the $KN$ and $\bar DN$ results we present them as a function of the
corresponding excess energies. (One should be aware that a comparison
at the same laboratory momentum, say, would look quite different because
of the large mass difference between the kaon and the $D$ meson.)
Within the range shown the $KN$ and $\bar DN$ reactions are predominantly
elastic. The first inelastic hadronic channel (pion production) opens
at an excess energy of around 136 MeV. We want to mention also that, due
to the much smaller mass difference between $D^*$(2010) and $D$(1869)
versus $K^*$(892) and $K$(496), the nominal threshold of the $D^*N$
channel occurs at a significantly smaller excess energy than the
corresponding $K^*N$ channel. Indeed the former practically coincides
with the $DN\pi$ threshold.

Based on the $\bar DN$ interaction potential ${\cal V}$ described in the two
preceeding sections the corresponding reaction amplitude ${\cal T}$ is
obtained by solving a Lippmann-Schwinger type scattering equation
defined by the time-ordered perturbation theory,
\begin{eqnarray}
\nonumber
{\cal T} = {\cal V} + {\cal V} {\cal G}_0 {\cal T} \ ,
\label{LSE}
\end{eqnarray}
from which we calculate the $\bar DN$ observables in the standard
way \cite{Juel1}. Due to the large mass of the $D$ meson that enters
into the free Green's function $G_0$ higher iterations play a
somewhat less important role for $\bar DN$ as compared to the $KN$ system.
But we want to emphasize that unitarization of the reaction
amplitude, which is achieved by solving Eq. (\ref{LSE}), is essential
for obtaining meaningful results because the resulting phase shifts
in the $S$- and also $P$-waves are in the order of 20 degrees or even
more in the energy range covered by our study. 
For completeness let us
also mention that we use averaged masses for the $D$ mesons, namely
$m_D$=1866.9 MeV and $m_{D^*}$=2009 MeV.

\begin{figure}
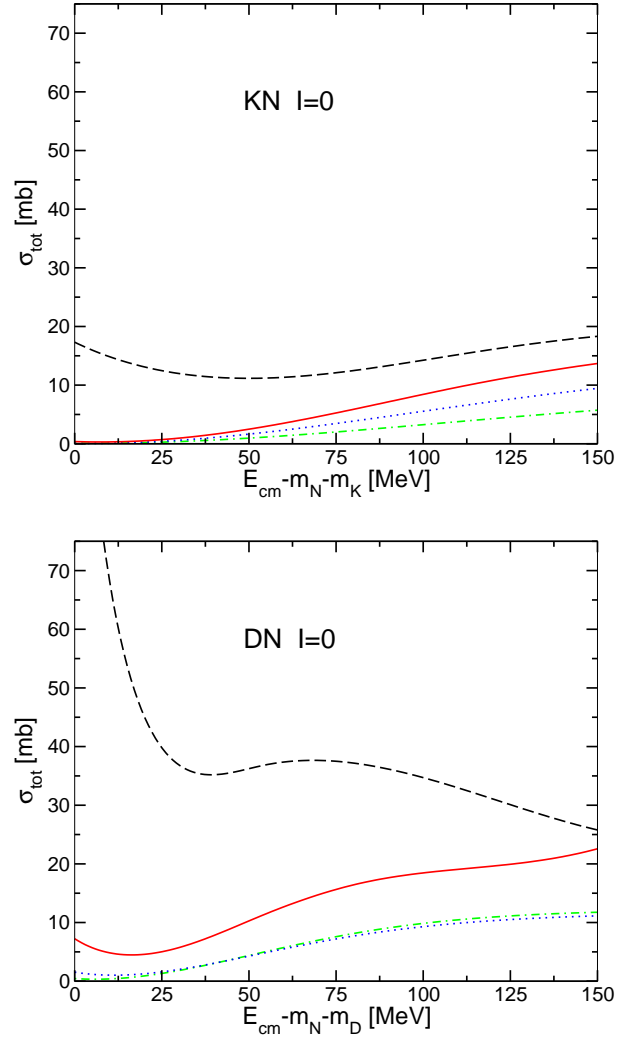

\epsfig{file=ksig0.eps, width=8.0cm}
\vskip 0.5cm
\epsfig{file=dsig0.eps, width=8.0cm}
\caption{$KN$ and $\bar D N$ cross sections in the isospin channel I=0
including consecutively $\rho$ (dashed curve), $\omega$ (dash-dotted),
scalar mesons and baryon-exchange diagrams (dotted),
and box diagrams (solid).
}
\label{DNM0}
\end{figure}

\begin{figure}
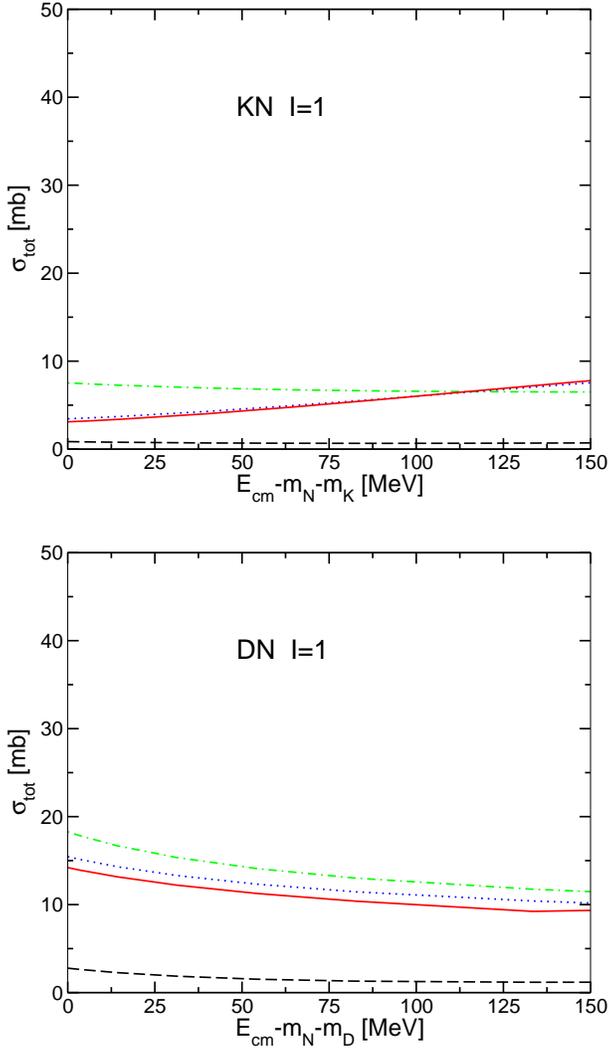

\epsfig{file=ksig1.eps, width=8.0cm}
\vskip 0.5cm
\epsfig{file=dsig1.eps, width=8.0cm}
\caption{$KN$ and $\bar D N$ cross sections in the isospin channel I=1.
Same description of curves as in Fig.~\ref{DNM0}.
}
\label{DNM1}
\end{figure}

Results based on the meson-exchange contributions are shown in
Figs.~\ref{DNM0} (for isospin $I=0$) and \ref{DNM1} (for $I=1$). Since
some of the $\bar DN$ model calculations in the literature take into
account only $\rho$ exchange \cite{Lin00,Sib01} we consider its
contribution first. It is obvious that the resulting cross
sections for the $\bar DN$ system are much larger than those for
$KN$. But one should keep in mind that this difference is primarily
caused by the different kinematics (masses). The involved coupling
constants are exactly the same for the $KN$ and $\bar DN$ interactions
under the assumption of SU(4) symmetry, as discussed in Sect. 2.
It is worth noting that the cross section in the $I=0$ channel
of the $\bar DN$ is particularly large and even exceeds 100 mb near
threshold, while for $I=1$ is is less then 5 mb over the whole
considered energy range.

The picture changes drastically once $\omega$ exchange is added.
Specifically, in the $I=0$ channel its contribution interferes
distructively with the (attractive) $\rho$ exchange and leads to a
strong reduction of the predicted cross section. On the other hand,
for $I=1$ both contributions are repulsive and add up
so that now the cross sections in both isospin channels are of
comparable magnitude. Indeed after
inclusion of the $\omega$ exchange the $KN$ results show already the
typical features of the full model \cite{Juel2} but also of the
experimental information \cite{Sib06}, namely an almost constant cross
section for $I=1$ and a cross section for $I=0$ that is practically
zero at threshold and then increases with energy. The predictions
for the $\bar DN$ system exhibit very similar features.

The addition of the scalar contributions and of baryon
($\Lambda_c$(2285), $\Sigma_c$(2455)) exchange influences the results
for $\bar DN$ very little and therefore we don't show them separately.
This is not too surprising in view of our assumption about the
origin of the scalar sector (we will come back to this issue later)
and of the large mass of the exchanged baryons. On the other hand,
the box diagrams yield a sizeable contribution, in particular in the
$I=0$ channel of the $\bar DN$ system.

\begin{figure}
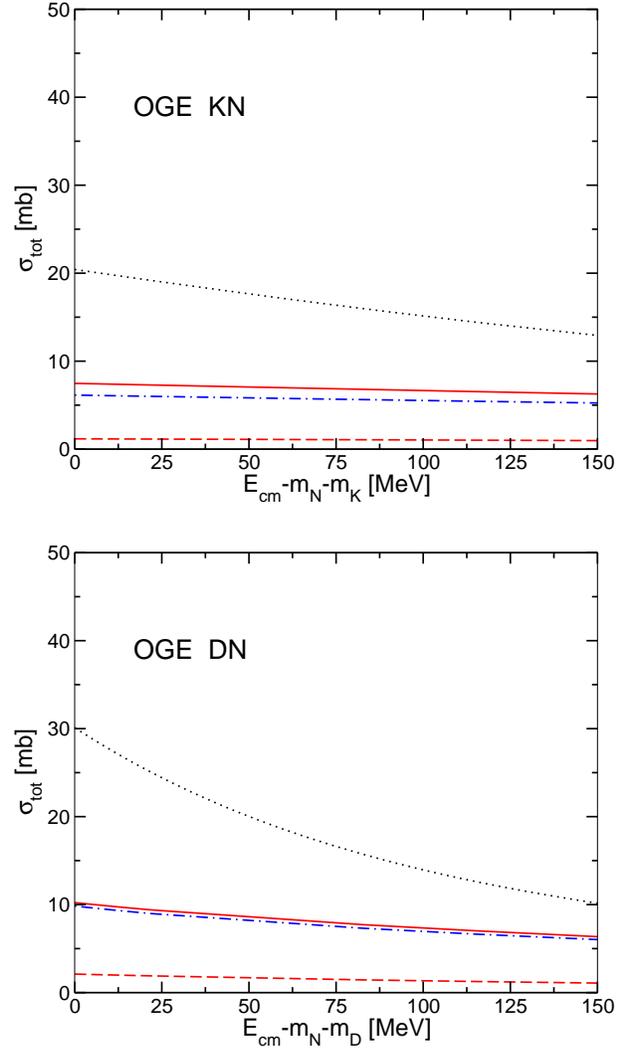

\epsfig{file=qsigk.eps, width=8.0cm} 
\vskip 0.5cm
\epsfig{file=qsigd.eps, width=8.0cm}
\caption{$KN$ and $\bar D N$ cross sections from OGE.
The dashed-dotted (dashed) curve is the result for the
spin-spin ($SS$) part in the $I=1$ ($I=0$) channel. 
The solid curve is the result for the $I=1$ channel after 
adding the Coulomb part. The dotted curve is the corresponding
cross section as obtained in Born approximation. 
For the $I=0$ channel the Coulomb component is zero so that
the full OGE result coincides with the curve for the $SS$ part.
}
\label{DNQ}
\end{figure}

Results for the $KN$ and $\bar DN$ interactions based on the OGE in the 
quark model are presented in Fig.~\ref{DNQ}. We use standard quark model
parameters~\cite{Barnes:1992ca}. For the light quark masses we take 
$m_u = m_d = 330$~MeV and for the strange and charm quark masses we 
use $m_s = 550$~MeV and $m_c = 1600$~MeV, and the quark-gluon 
coupling is taken to be $\alpha_s = 0.6$. The size parameters of 
the nucleon and the kaon wave functions are $\alpha_N = 400$~MeV and 
$\beta_D = 350$~MeV. For the $D$ meson wave function we use the value
of Ref.~\cite{Martins:1994hd}, namely $\beta_D = 383.5$~MeV. The results 
demonstrate that, as in the case of $KN$, the spin-spin component of the OGE 
is much more important than the Coulomb component. 
(Note that the contribution of the Coulomb component is zero in the 
$I=0$ channel.) 
We have performed exploratory calculations utilizing a larger value for 
$\beta_D$ for the $D$ meson, $\beta_D = 440$~MeV, as given by a recent 
calculation~\cite{Hilbert:2007hc}. The results for the combined Coulomb 
and spin-spin OGE do not change appreciably, although the Coulomb part is a 
little smaller in this case as compared to the corresponding value with 
$\beta_D = 383.5$~MeV.

The cross sections predicted for the $\bar DN$ system are roughly a factor 
1.5 larger than those for $KN$. Note that, unlike for the meson-exchange part, 
here the parameters entering the potential differ, reflecting the different 
quark masses and sizes of the $K$ and $\bar D$ mesons. 
As in the $KN$ case, graphs (2) and (4) in 
Fig.~\ref{fig:interch} are suppressed as compared to graphs (1) and (3), because 
of the large quark mass in the denominator of Eq.~(\ref{VqS}). 
This suppression is even stronger in the $\bar DN$ case. Note that graphs (1) and 
(3) actually become somewhat larger as compared to the kaon case because of the 
larger size of the $D$ meson. 
 
It is worth pointing out there are sizable effects due to the iteration 
of the interaction in the Lippmann-Schwinger equation. In order to
demonstrate this we include also results obtained in Born approximation 
for the $I=1$ channel. These are shown by the dotted lines in Fig.~\ref{DNQ}. 
Obviously, close to threshold the Born result for the $\bar DN$ cross section is 
of the order of $30$~mb, while the corresponding unitarized result is of the 
order of $10$~mb, i.e. there is more than 50~\% difference. 

\begin{figure}
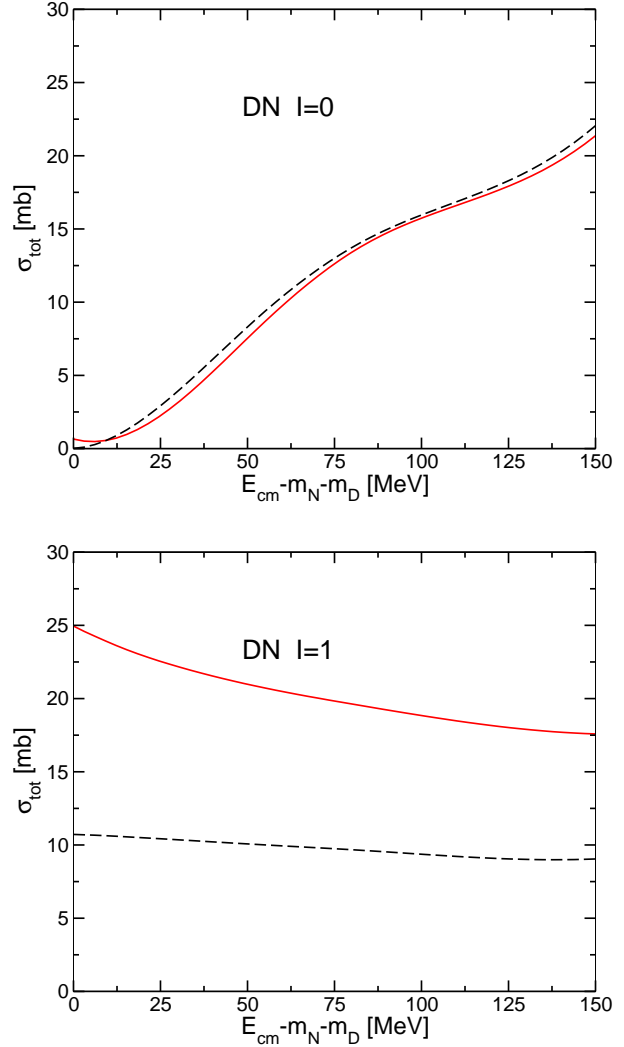

\epsfig{file=sig0.eps, width=8.0cm}
\vskip 0.5cm
\epsfig{file=sig1.eps, width=8.0cm}
\caption{$\bar D N$ cross sections in the isospin channels I=0,1.
The solid curves are the results of the full model, i.e. including
meson-exchange and OGE, and viewing the scalar contributions to 
be due to correlated $\pi\pi$ exchange. The dashed curves show 
results obtained in the scenario that
attempts to simulate the case that the scalar contributions are 
due to genuine scalar-meson exchange, cf. text. 
}
\label{DNF}
\end{figure}

Our full predictions for the $\bar DN$ cross sections, combining
now the mesonic part with the contributions from quark-gluon
processes, are presented in Fig.~\ref{DNF}. The solid 
lines are results for the scenario where the scalar-meson contributions
are viewed as being due to correlated $\pi\pi - K\bar K$ exchange,
in line with the philosophy of the original J\"ulich $KN$ model
\cite{Juel2}. In this case the cross section for $I=1$ is in
the order of 20 mb, i.e. roughly twice as large as observed for
the $KN$ system \cite{Sib06}. For the $I=0$ channel we predict
a cross section that is practically zero at the threshold but
increases to about 25~~mb at the excess energy 150 MeV. Also
here the result is roughly twice as large as the cross section
for $KN$ at the corresponding excess energy.

\begin{figure}
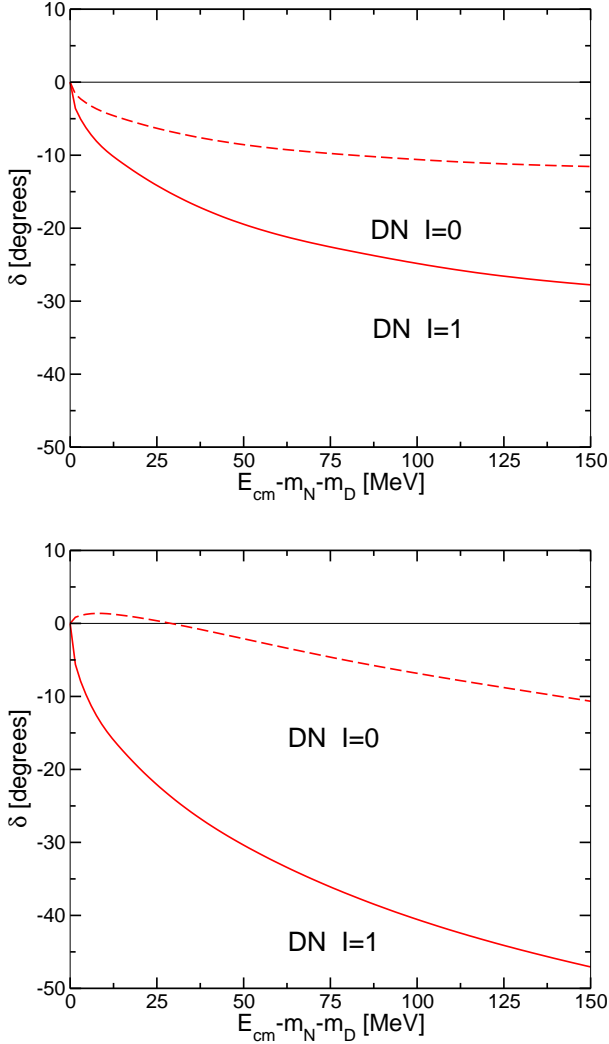

\epsfig{file=phasq.eps, width=8.0cm}
\vskip 0.5cm
\epsfig{file=phasm.eps, width=8.0cm}
\caption{$\bar D N$ $S$-wave phase shifts in the isospin channels I=0,1.
The upper panel shows the results for the quark model based on OGE 
while the lower pannel are the results of the full model, i.e. 
including meson-exchange and OGE, and viewing the scalar contributions 
to be due to correlated $\pi\pi$ exchange. 
}
\label{DNPh}
\end{figure}

The dashed curves show results obtained in the scenario that
attempts to simulate the case that the scalar contributions are 
due to genuine scalar-meson exchange. In order
to get a rough estimate for that scenario we assume here, for
simplicity reasons, that the couplings for $\bar D N$
are the same as for $KN$ for the two scalar mesons in question.
A strict evaluation within the SU(4)
scheme would involve several unknown quantities such as the
singlet couplings and the mixing angles and is not feasible.
In any case, our prescription is only meant to illustrate the
ambiguity resulting from the unclear situation in the scalar sector.
Evidently, the cross section for $I=0$ is rather insensitive to the
treatment of the scalar mesons, at least within the scenarios considered
here. Indeed, due to their isospin structure, the contributions of the
$\sigma$ and the $a_0$(980) mesons tend to cancel in this channel.
On the other hand, the predicted $I=1$ cross section reduces by
50~\% for the scenario based on larger scalar-meson coupling constants.
This variation may be considered as a measure for the uncertainty in our
model prediction for $\bar D N$, despite of constructing the interaction
in close analogy to $KN$ and invoking strict SU(4) symmetry.

Finally, for completeness let us also present the $S$-wave $\bar DN$ 
phase shifts. Corresponding results are shown in Fig.~\ref{DNPh}, for the
quark-gluon interaction alone (upper panel) and for the full model 
(lower panel). Obviously, in general all interactions are repulsive, 
as reflected in the negative sign of the phase shifts. But the 
full model is weakly attractive for energies near the threshold in
the $I=0$ channel. 
The corresponding scattering lengths are 
$a^{I=0}= -0.13\,$fm, $a^{I=1}= -0.29\,$fm, 
for the quark-gluon interaction and 
$a^{I=0}= 0.07\,$fm, $a^{I=1}= -0.45\,$fm,
for the full model. Interestingly, the former results are pretty close to
the values found by Lutz and Korpa for their $\bar DN$ interaction 
\cite{Lutz05} while the latter are qualitatively very similar to the 
results obtained for the $KN$ interaction \cite{Juel2}.

\section{Summary}
\label{sec:sum}

In this paper we presented predictions for the low-energy $\bar DN$
cross section based on a model which was developed in close analogy to
the meson-exchange $KN$ interaction of the J\"ulich group \cite{Juel1,Juel2},
utilizing SU(4) symmetry constraints. The main ingredients of
the interaction are provided by vector meson ($\rho$, $\omega$)
exchange but higher-order box diagrams involving ${\bar D}^*N$, 
$\bar D\Delta$, and ${\bar D}^*\Delta$ intermediate states, are taken into 
account too. Furthermore, in the spirit of a recent study of the $KN$ system by us
\cite{HHK}, the short range part is again assumed to receive additional
contributions from genuine quark-gluon processes.

The cross sections for $\bar DN$ predicted for excess energies up to
150 MeV are of the same order of magnitude as those
for $KN$ but with average values of around 20~mb roughly a factor two
larger than for the latter system. There is an uncertainty in our
prediction for the $I=1$ channel which is caused by the unknown SU(4)
structure of the scalar-meson sector. Assuming that the contributions
in the scalar sector are due to correlated $\pi\pi$ exchange, in
line with the J\"ulich $KN$ model, we find that the scalar
contributions influence the $\bar DN$ cross sections only 
marginally. 
But a scenario where the effect of the exchange of genuine scalar
mesons is simulated by assuming their coupling strengths to be the 
same as in the $KN$ model 
leads to a 50 \% reduction of the $I=1$ cross section. 

Anyway, the most interesting finding of our study is certainly the
important role played by the $\omega$-exchange contribution. Its
interference pattern with the $\rho$-exchange, which is basically
fixed by the assumed SU(4) symmetry, clearly determines the qualitative
features of the $\bar DN$ interaction -- similiar to what happens also for 
the $KN$ system. As a consequence also the cross sections predicted for $\bar DN$ 
show qualitatively very similar features to those known of $KN$ scattering.
On the other hand, predictions for $\bar DN$ where only $\rho$-exchange
was taken into account differ drastically and, in our opinon, should be
regarded with caution in view of the results presented in this paper.

\subsection*{Acknowledgements}
This work was financially supported by the
Deutsche For\-schungsgemeinschaft (Project no. 444 BRA-113/14) and
the Brazilian agencies CAPES, CNPq and FAPESP.
This work was supported in part by the EU I3HP
\lq\lq Study of Strongly Interacting Matter'' under contract number
RII3-CT-2004-506078, by the DFG through funds provided
to the SFB/\-TR 16 \lq\lq Subnuclear Structure of Matter'',
by the  EU Contract No. MRTN-CT-2006-035482, \lq\lq FLAVIAnet'' 
and by  BMBF (grant 06BN411). A.S. acknowledges support by the
JLab grant SURA-06-C0452  and the COSY FFE grant No. 41760632 (COSY-085).

\vfill\eject
\appendix

\section{The interaction Lagrangians}
\label{app:lags}

Here we list the specific interaction Lagrangians which are used to
derived the meson-exchange $\bar DN$ interaction. The
baryon-baryon-meson couplings are given by
\bea
\nonumber
{\cal L}_{NNS} &=& g_{NNS} \bar \Psi_N(x) \Psi_N(x) \Phi_S(x) \ , \\
\nonumber
{\cal L}_{NNP} &=& g_{NNP} \bar \Psi_N(x) i\gamma^5 \Psi_N(x) \Phi_P(x) \ , \\
\nonumber
{\cal L}_{NNV} &=& g_{NNV} \bar \Psi_N(x) \gamma_\mu \Psi_N(x) \Phi_V^\mu (x) \\
\nonumber
&&+
\frac{f_{NNV}}{4 m_N} \bar \Psi_N(x) \sigma_{\mu\nu} \Psi_N(x)
(\partial^\mu \Phi_V^\nu (x) - \partial^\nu \Phi_V^\mu (x)) \ , \\
\nonumber
{\cal L}_{N\Delta P} &=& \frac{f_{N\Delta P}}{m_P} \bar \Psi_{\Delta \mu}(x) \Psi_N(x)
\partial^\mu \Phi_P (x) + H.c. \ , \\
\nonumber
{\cal L}_{N\Delta V} &=& \frac{f_{N\Delta V}}{m_V} i (\bar \Psi_{\Delta \mu}(x) 
\gamma^5 \gamma_\mu \Psi_N(x) \\
\nonumber
&&- \bar \Psi_N(x) \gamma^5 \gamma_\mu \Psi_{\Delta \mu}(x))
(\partial^\mu \Phi_V^\nu (x) - \partial^\nu \Phi_V^\mu (x)) \ , \\
\nonumber
{\cal L}_{N Y P} &=& \frac {f_{N Y P}}{m_P} 
 (\bar \Psi_{Y}(x) \gamma^5 \gamma^\mu \Psi_N\,(x) \\ 
&&+ \bar \Psi_N(x) \gamma^5 \gamma^\mu \Psi_{Y}\,(x)) \partial_\mu 
     \phi_P (x) \; .
\eea
Here $\Psi_N$, $\Psi_{\Delta\mu}$, and $\Psi_Y$ are the nucleon, $\Delta$, and hyperon
field operators and $\Phi_S$, $\Phi_P$, and $\Phi_V^\mu$ are the field operators for scalar,
pseudoscalar and vector mesons, respectively.

The employed three-meson couplings are
\begin{eqnarray}
\nonumber
{\cal L}_{PPS} &=& g_{PPS} m_P \Phi_P(x) \Phi_P(x) \Phi_S(x) \ , \\
\nonumber
{\cal L}_{PPV} &=& g_{PPV} \Phi_P(x) \partial_{\mu} \Phi_P(x) \Phi_V^{\mu}(x) \ , \\
\nonumber
{\cal L}_{VVP} &=& \frac{g_{VVP}}{m_V}i\epsilon_{\mu \nu \tau \delta}
\partial^\mu \Phi_V^\nu (x) \partial^\tau \Phi_V^\delta (x) \Phi_P(x) \ , \\
&&
\end{eqnarray}
where $\epsilon_{\mu \nu \tau \delta}$ is the antisymmetric tensor with
$\epsilon^{0123}=1$. Note that here only the space-spin part is given. The
additional SU(4) flavour structure that leads to the characteristic relations
between the coupling constants is discussed in Sect. II. Details on the
derivation of the meson-baryon interaction potential from those Lagrangians
can be found in Refs. \cite{Juel1,Juel2}.

\section{The quark-model meson-baryon interaction}
\label{app:qm}

In this Appendix we outline the derivation of the effective
meson-baryon interaction in the quark model. As already mentioned, 
the effective interaction is given by the quark-Born diagrams
depicted in Fig.~\ref{fig:interch}. Their expressions can be
obtained in the Born-order quark interchange model~\cite{Barnes:1992qa},
or using the methods of the resonating group, or the Fock-Tani 
representation~\cite{Hadjimichef:1998rx}. Given the interactions and the 
bound state amplitudes of the single hadrons, the expression of the quark 
Born diagrams for the effective meson-baryon interaction 
$\alpha + \beta \rightarrow \gamma + \delta$ is given by
\bea 
\nonumber 
{\cal V}_{MB}^{(\alpha\beta;\gamma\delta)} 
= &-& 3 \, \Phi^{*\mu
\nu_1}_\gamma \, \Psi^{*\nu\mu_2\mu_3}_\delta \,
V_{qq}(\mu\nu;\sigma\rho) \, \Phi^{\rho\nu_1}_\alpha \,
\Psi^{\sigma\mu_2\mu_3}_\beta  \\
\nonumber 
&-& 3 \, \Phi^{*\sigma\rho}_\gamma \, \Psi^{*\mu_1\mu_2\mu_3}_\delta
\, V_{q\bar{q}}(\mu\nu;\sigma\rho) \, \Phi^{\mu_1\nu}_\alpha \,
\Psi^{\mu\mu_2\mu_3}_\beta \\
\nonumber 
&-& 6 \, \Phi^{*\mu_1 \nu_1}_\gamma \, \Psi^{*\nu\mu\mu_3}_\delta \,
V_{qq}(\mu\nu;\sigma\rho) \, \Phi^{\rho\nu_1}_\alpha \,
\Psi^{\mu_1\sigma\mu_3}_\beta \\
\nonumber 
&-& 6 \, \Phi^{*\mu_1 \nu}_\gamma \, \Psi^{*\nu_1\mu\mu_3}_\delta \,
V_{q\bar{q}}(\mu\nu;\sigma\rho) \,\Phi^{\nu_1\rho}_\alpha \,
\Psi^{\mu_1\sigma\mu_3}_\beta \label{VMB}. \\
& &
\eea
Here, the $\Phi$ and $\Psi$ are Fock-space amplitudes of the
one-meson and one-baryon states, which in a second quantization
notation are given as
\beq |M_\alpha\rangle = \Phi^{\mu\nu}_\alpha \, q^\dag_\mu \, {\bar
q}^\dag_\nu |0 \rangle, \ |B_\alpha\rangle  =
\frac{1}{\sqrt{3!}} \, \Psi^{\mu_1\mu_2\mu_3}_\alpha \,
q^\dag_{\mu_1} \, q^\dag_{\mu_2} \, q^\dag_{\mu_3}|0\rangle ,
\label{mes-bar} \eeq
where $\alpha$ indicates all quantum numbers necessary to specify
the hadronic state, like c.m. momentum, spin and flavor, and $\mu,
\nu, \cdots$ indicate all quantum numbers of the quarks like
momentum, color, spin and flavor -- a sum or integral over repeated
indices is implied. $q^\dag$, $\bar{q}^\dag$, $q^\dag$, and
$\bar{q}$ are quark and antiquark creation and anhihilation
operators that satisfy the usual canonical anticommutation
relations. In addition, $V_{qq}$, $V_{\bar{q}q}$ and
$V_{\bar{q}\bar{q}}$ are the microscopic quark and antiquark
interactions, which in the same second quantization notation are
defined through
\bea
V &=& \frac{1}{2} V_{qq}(\mu\nu;\rho\sigma) q^\dag_\mu \,
q^\dag_\nu \, q_\sigma \, q_\rho + \frac{1}{2}
V_{\bar{q}\bar{q}}(\mu\nu,\rho\sigma) \bar{q}^\dag_\mu \,
\bar{q}^\dag_\nu  \,  \bar{q}_\sigma  \, \bar{q}_\rho \nn
\\
&+& V_{q\bar{q}}(\mu\nu;\rho\sigma) q^\dag_\mu  \, \bar{q}^\dag_\nu
\, \bar{q}^\dag_\sigma  \,  q^\dag_\rho.
\label{V}
\eea
The expression for ${\cal V}_{MB}$ in Eq.~(\ref{VMB}) involves a
6-dimen\-sional integral that cannot be integrated analytically for
general forms of the amplitudes $\Phi$ and $\Psi$. However, when
using Gaussian forms for the meson and baryon amplitudes, many of
the integrals can be done analytically and the resulting expression
for each of the diagrams of Fig.~\ref{fig:interch} is of the form
given in Eq.~(\ref{Vi}). Specifically, we use for the amplitudes
$\Phi$ and $\Psi$ in momentum space Gaussian forms with width
parameters $\beta_D$ and $\alpha_N$ as
\beq \Phi_{\vec{P}}(\vec{k}_1,\vec{k}_2) =
\delta^{(3)}\Bigl(\vec{P}-\vec{k}_1 - \vec{k}_2\Bigr)
\left(\frac{1}{\pi \beta^2_D}\right)^{3/2} e^{-
k^{\,2}_{rel}/8\beta^2_D} , \eeq
where
\beq
\vec{k}_{rel} = \frac{2(M_{\bar{q}}\vec{k}_1 - M_q
\vec{k}_2)}{M_q + M_{\bar{q}}},
\eeq
and
\bea 
\nonumber
\Psi_{\vec{P}}(\vec{k}_1,\vec{k}_2,\vec{k}_3) =&&
\delta^{(3)}\Bigl(\vec{P}-\sum^3_{i=1} \vec{k}_i\Bigr)
\left(\frac{3}{\pi^2\alpha^4_N}\right)^{3/4} \\ 
&&\times e^{- \sum^3_{i=1}
(\vec{k}_i-\vec{P}/3)^2/2\alpha^2_N} . 
\eea
Writing
\beq V_{qq}(\mu\nu;\sigma\rho) = \delta(\vec{k}_\mu + \vec{k}_\nu -
\vec{k}_\sigma - \vec{k}_\rho) \, v(\vec{k}_\mu - \vec{k}_\rho),
\eeq
and equivalently for $V_{\bar{q}q}$ and $V_{\bar{q}\bar{q}}$,
after integrating over the quark momenta as indicated in
Eq.~(\ref{VMB}) one obtains the expression given in Eq.~(\ref{Vi}),
where the $a_i, b_i, \cdots,$ can be written as a ratio $a_i =
n(a_i)/d(a_i)\alpha^2_N, b_i=n(b_i)/d(b_i)\alpha^2_N$, etc. The
corresponding expressions are given as follows:

\noindent Graph (1):
\bea
n(a_1) &=&  3\,g^2 + 3\,{\left( 1 + \rho \right) }^2
+ g\,\left( 7 + 8\,\rho + 10\,{\rho}^2 \right)  \nn \\
d(a_1) &=&  {6\,{\left( 3 + 2\,g \right)
\,{\left( 1 + \rho \right) }^2 }} \nn \\
n(c_1) &=&  g^2 + {\left( 1 + \rho \right)}^2
- 2\,g\,\left( -1 + {\rho}^2 \right)  \nn \\
d(c_1) &=& {\left( 3 + 2\,g \right) \,{\left( 1 + \rho \right) }^2}
\nn \\
n(d_1) &=& 3\,g, \hspace{0.5cm} d(a_1) = \left( 3 + 2\,g \right) \nn \\
n(e_1) &=& - g\, \,\left( 1 + 4\,\rho \right)\left( \vec{p} +
\vec{p}\,' \right)
\nn \\
d(e_1) &=& {\left( 3 + 2\,g \right) \,\left( 1 + \rho \right)},
\eea

\noindent Graph (2):
\bea
n(a_2) &=& n(a_1), \hspace{0.5cm}  d(a_2) = d(a_1) \nn \\
n(b_2) &=& n(b_1), \hspace{0.5cm}  d(b_2) = d(b_1) \nn \\
n(c_2) &=& n(c_1), \hspace{0.5cm}  d(c_2) = d(c_1) \nn \\
n(d_2) &=& {g\,\left( 3 + g \right) }, \hspace{0.5cm} d(d_2) =
{2\,\left( 3 + 2\,g \right)} \nn \\
n(e_2) &=& {g\,\left[\left( 2 + g + 2\,\rho \right) \vec{p} -
\left(1 + g - 2\,\rho \right) \vec{p}\,'\right] } \nn \\
d(e_2) &=& {\left( 3 + 2\,g \right) \, \left( 1 + \rho \right)},
\eea

\vspace{0.25cm} \noindent Graph (3):
\bea
n(a_3) &=& 3 g^2 + 3 \left(1+\rho\right)^2 + g \left(7 + 8 \rho + 10 \rho^2\right)  \nn \\
d(a_3) &=& {6\,\left( 3 + 2\,g \right) \, {\left( 1 + \rho \right) }^2} \nn \\
n(b_3) &=& n(a_3), \hspace{0.5cm} d(b_3) = d(a_3)  \nn \\
n(c_3) &=& n(c_1), \hspace{0.5cm} d(c_3) = d(c_1) \nn \\
n(d_3) &=& {6 + 7\,g}, \hspace{0.5cm} d(d_3) = {4\,\left( 3 + 2\,g
\right)} \nn \\
n(e_3) &=& -3 \left[1 + g + \left(1 + 2 g\right) \rho \right]
\vec{p} + \left[3 + g + \left( 3 - 2 g \right) \rho\right] \vec{p}\,' \nn \\
d(e_3) &=& {2\,\left( 3 + 2\,g \right) \,\left( 1 + \rho \right) },
\eea

\vspace{0.25cm} \noindent Graph (4):
\bea
n(a_4) &=& n(a_1), \hspace{0.5cm} n(b_4) = n(b_1) \nn \\
d(a_4) &=& d(a_1), \hspace{0.5cm} d(b_4) = d(b_1) \nn \\
n(c_4) &=& n(c_1), \hspace{0.5cm} d(c_4) = d(c_1) \nn \\
n(d_4) &=& {2 + g}, \hspace{0.5cm} d(d_4) = {4} \nn \\
n(e_4) &=& - {\left( 1 + g + \rho \right)\, \left( \vec{p} -
\vec{p}\,'
\right)} \nn \\
d(e_4) &=& {2\,\left( 1 + \rho \right)}.
\eea
In these, $g=\alpha^2_N/\beta^2_D$ and $\rho = M_u/M_c$.

\end{document}